\documentclass[12pt]{iopart}

\usepackage[english]{babel}
\usepackage[utf8x]{inputenc}
\usepackage[T1]{fontenc}

\usepackage{iopams}
\usepackage{bm}
\usepackage{xcolor}
\usepackage[colorlinks,allcolors=black]{hyperref}
\usepackage{graphicx}
\usepackage{dsfont}
\usepackage{booktabs}
\usepackage[acronym,nohypertypes={acronym}]{glossaries-prefix}
\usepackage{relsize}
\usepackage{placeins}
\usepackage[print-unity-mantissa=false]{siunitx}
\usepackage[justification=justified]{caption}
\usepackage{mathtools}

\newacronym[longplural=amplitude spectral densities]{asd}{ASD}{amplitude spectral density}

\newacronym{cic}{CIC}{cascaded integrator-comb}

\newacronym{esa}{ESA}{European Space Agency}
\newacronym{emris}{EMRIs}{extreme mass-ratio inspirals }

\newacronym{fir}{FIR}{finite impulse response}

\newacronym{gw}{GW}{gravitational wave}

\newacronym[prefixfirst={the\ }]{lisa}{LISA}{Laser Interferometer Space Antenna}
\newacronym{lot}{LOT}{LISA On Table}
\newacronym{lti}{LTI}{linear time-invariant}

\newacronym[longplural=movable optical sub-assemblies]{mosa}{MOSA}{movable optical sub-assembly}

\newacronym{rfi}{rfi}{reference interferometer}

\newacronym[longplural=power spectral densities]{psd}{PSD}{power spectral density}

\newacronym{isi}{isi}{inter-spacecraft interferometer}
\newacronym{sc}{SC}{spacecraft}

\newacronym{tdi}{TDI}{time-delay interferometry}
\newacronym{tmi}{tmi}{test-mass interferometer}

\newcommand{\decimate}[0]{\mathbf{S}}
\newcommand{\filter}[0]{\mathbf{F}}
\newcommand{\dfilter}[0]{\mathbf{G}}
\newcommand{\delay}[0]{\mathbf{D}}
\newcommand{\fdelay}[0]{\mathcal{D}}
\newcommand{\tdi}[0]{\mathbf{C}}
\newcommand{\diff}[0]{\mathbf{\Delta}}

\DeclareMathOperator{\rect}{rect}

\begin{document}

\title[Mitigating Flexing-Filtering]{Mitigation of the Flexing-Filtering Effect in Time-Delay Interferometry}

\author{Shivani Harer$^1$, Martin Staab$^{2,3}$, Hubert Halloin$^1$}

\address{$^1$ Universit\'e Paris Cit\'e, CNRS, Astroparticule et Cosmologie, 10 rue Alice Domon et L\'eonie Duquet, 75013 Paris, France}
\address{$^2$ LTE, Observatoire de Paris, Universit\'e PSL, Sorbonne Universit\'e, Universit\'e de Lille, LNE, CNRS 61 Avenue de l’Observatoire, 75014 Paris, France}
\address{$^3$ Institute for Gravitational and Subatomic Physics (GRASP), Department of Physics, Utrecht University, Princetonplein 1, NL-3584 CC Utrecht, The Netherlands}

\ead{harer@apc.in2p3.fr}
\vspace{10pt}
\begin{indented}
\item[]September 2025
\end{indented}

\begin{abstract}

In early 2024, ESA formally adopted the Laser Interferometer Space Antenna (LISA) space mission with the aim of measuring gravitational waves emitted in the millihertz range. The constellation employs three spacecraft that exchange laser beams to form interferometric measurements over a distance of 2.5 million kilometers. The measurements will then be telemetered down to Earth at a lower sampling frequency. Anti-aliasing filters will be used on board to limit spectral folding of out-of-band laser noise. The dominant noise in these measurements is laser frequency noise which does not cancel naturally in LISA's unequal-arm heterodyne interferometers. Suppression of this noise requires time-shifting of the data using delay operators to build  virtual beam paths that simulate equal-arm interferometers. The non-commutativity of these delay operators and on-board filters manifests as a noise (flexing-filtering) that significantly contributes to the noise budget. This non-commutativity is a consequence of the non-flatness of the filter in-band. Attenuation of this noise requires high-order and computationally expensive filters, putting additional demands on the spacecraft. The following work studies an alternative method to reduce this flexing filtering noise via the introduction of a modified delay operator that accounts for the non-commutativity with the filter in the delay operation itself. Our approach allows us to reduce the flexing-filtering noise by over six orders of magnitude whilst reducing the dependency on the flatness of the filter.  The work is supplemented by numerical simulations of the data processing chain that compare the results with those of the standard approach. 
\end{abstract}

\submitto{\CQG}
\noindent{\it Keywords}: LISA, gravitational-wave detection, time-delay interferometry, laser-noise suppression, flexing-filtering effect 

\clearpage

\glsresetall
\section{Introduction}
\label{sec:intro}
\Pgls{lisa} mission is a future, space-based \gls{gw} observatory, that will employ interferometric techniques to detect the waves with a sub-picometre precision. The mission is being led by the \gls{esa} and is scheduled to launch in the year 2035. The observatory will be sensitive to GW frequencies between \qty{0.1}{\milli\Hz} and \qty{1}{\Hz}, emitted during several large-scale galactic events. This includes but is not limited to quasi-monochromatic sources like galactic binaries, those emitted by the coalescences of extreme mass-ratio inspirals and by the merging of massive Black Hole binaries. Studying these events will not only allow us to trace origins and histories of the objects themselves, but also study the structure and rate of expansion of the Milky Way galaxy, explore the fundamental nature of gravity and explore the stochastic GW background~\cite{LISA:2024hlh}.

The LISA constellation design consists of three \gls{sc} in a triangle formation, conducting cartwheel motion in a heliocentric orbit placed around \qty{20}{\degree} behind the Earth. These spacecraft will exchange laser beams over a free-space distance of 2.5 million \unit{\kilo\m}, implementing laser interferometry to attain a strain sensitivity of the order of \num{e-21} to \num{e-23}~\cite{LISA:2024hlh}. The three spacecraft will be identical, each carrying two \glspl{mosa}, housing an optical bench with a laser system, a free-falling test mass and a phase measurement system. The \glspl{mosa} exchange laser light via backlink fibers and telescopes to monitor the differential laser phase. LISA will employ the technique of split-interferometry, i.e. measurements will be made by three different interferometers on the optical bench; the inter-spacecraft interferometer, the reference interferometer and the test mass interferometer. The inter-spacecraft interferometer combines laser light from the distant and local lasers and will therefore contain the \gls{gw} signal. The reference interferometer compares the local laser with that of adjacent MOSA on the same spacecraft to serve as a measure of relative phase between the two. The local laser is bounced off the test-mass before recombination with the laser of the adjacent MOSA in the test mass interferometer to quantify the longitudinal motion of the spacecraft with respect to the test mass~\cite{LISA:2024hlh}. The phasemeters aboard the spacecraft record the phase of the interferometric beatnotes at a sampling rate of \qty{80}{\mega\Hz}. This data must then be decimated to \qty{4}{Hz} before it is transmitted to Earth due to limitations in the telemetry data budget\footnote{Note that this decimation does not cause any data loss, since the frequencies in this range are dominated by instrumental noise and fall outside the \gls{lisa} detection band of \qty{0.1}{\milli\Hz} to \qty{1}{\Hz}.}. 
Using adequate anti-aliasing filters with high attenuation in the stop band~\cite{Staab:2023qrb} is crucial to limit folding of out-of-band laser noise power in the decimation operation~\cite{Schwarze:2018lvl,Vidal:inprogress}. Note that aliased noise cannot be mitigated by post-processing techniques. Therefore, sufficient stop-band attenuation is the main driver for the design of the anti-aliasing filters. 

\Pgls{lisa}'s ability to detect \glspl{gw} is subject to the sufficient reduction of various sources of noise. The primary source of noise for LISA is laser frequency noise. In ground-based detectors that apply equal-arm interferometry, this noise is subtracted upon recombination of the laser beams. In space however, the inter-spacecraft distances in will vary, a motion called \textit{flexing}, and result in the Doppler shifting of the laser received by the spacecraft. \Gls{tdi} is a post-processing technique that synthesizes virtual equal-arm interferometers by delaying and linearly combining the output of the split-interferometry configuration~\cite{Tinto:1999yr,Estabrook:2000ef,Tinto:2020fcc}. The technique of \gls{tdi} has been extensively studied and experimentally verified not only for LISA but also other space-based observatories~\cite{TianQin:2015yph,Hu:2017mde}. 

The post-processing step of \gls{tdi} is susceptible to additional noise due to the non-commutativity of its time-varying delay operation with the on-board processing steps of filtering and decimation~\cite{Staab:2023qrb}. The coupling between the flexing of the constellation and the anti-aliasing filters \textemdash coined \textit{flexing-filtering}~\cite{Bayle:2018hnm}\textemdash further reduces the efficiency of the \gls{tdi} algorithm. This coupling is proportional to the time-derivative of the delay and dependent on the choice of the anti-aliasing filter, specifically on the flatness of the filter response in the pass-band. Improving the flatness, however, comes with higher computational cost which puts additional demand on the spacecraft. Therefore, filter designs that fulfill some allotted requirement but use a minimum amount of computational resources are desirable.

Alternative methods have been suggested to minimize the coupling of the flexing-filtering effect. The potential use of a quasi-inverse filter in ground-processing pipelines to \textit{compensate} for the non-flat response of the anti-aliasing filter has been successfully implemented on simulated LISA data~\cite{Staab2023}. The application of such a filter to reduce the effect of flexing-filtering has also been discussed in the context of TDIR in the Hexagon experiment~\cite{Yamamoto:2024rgn}. However, the addition of more filters in the processing chain results in longer group delays in the interferometric data.

In this paper we propose a modified delay operation \textemdash that includes the quasi-inverse filter\textemdash as an alternative to the current delay operation used to time-shift signals for TDI. This approach sufficiently mitigates the flexing-filtering effect and the combination of the two individual operations will reduce computational load on the system caused by the additive nature of individually applied filters. We begin with the study of three different TDI approaches and their noise contributions in \sref{sec:tdi}. \Sref{sec:derivation} takes a closer look at the construction of the proposed modified delay operator and the derivation of its frequency response. In \sref{sec:residuals}, we compare the noise contributions to the three TDI approaches against the noise reference for LISA. The numerical implementation of the three approaches is presented in \sref{sec:Results}. The concluding remarks are discussed in \sref{sec:conclusions}. The source code of the algorithms presented in this paper including all datasets and analysis scripts are openly available~\cite{dataset}.

\section{Onboard Processing}
\label{sec:onboard}

\begin{figure}
    \includegraphics[width=\textwidth]{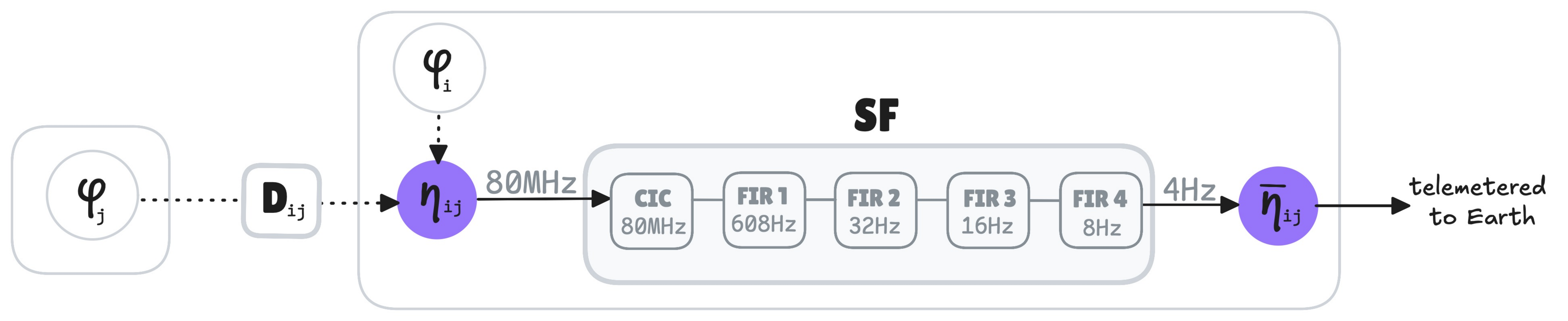} 
    \caption{A diagrammatic representation of the formation and processing of the beatnote phase $\phi_{\text{sc}}$ of the inter-spacecraft interferometer. The signal undergoes several steps of decimation using \acrfull{cic} and \acrfull{fir} filters before it is telemetered to Earth, where the data is further processed using \gls{tdi}.}
    \label{fig:onboard_processing}
\end{figure}

The building blocks for \gls{tdi} are the six one-way measurements constructed by interfering distant beams with local beams. In this section, we formulate these single laser-link measurements and explore their path before they are transmitted down to Earth. We simplify the description by skipping the subtraction of spacecraft motion~\cite{Estabrook:2000ef} and the reduction to three lasers. Figure \ref{fig:onboard_processing} shows a condensed schematic of the on-board processing circuit. A single-arm interferometric phase measurement between a distant spacecraft $j$ and local spacecraft $i$ is represented by
\begin{equation}
\label{eq:eta}
    \eta_{ij} = \mathbf{D}_{ij}\phi_j - \phi_i, 
\end{equation}
which is taken on spacecraft $i$ and compares the local laser phase $\phi_i$\footnote{The laser phase $\phi_i$ belongs to the laser source hosted on the right-hand side \gls{mosa}. A single index suffices here since the left-handed lasers are already canceled out.} with the distant laser phase $\phi_j$ which is emitted by the far spacecraft $j$. Here, $\mathbf{D}_{ij}$ denotes the delay operator that models the propagation of the laser phase from spacecraft $j$ to $i$. Its action on the distant laser phase $\phi_{j}(t)$ is defined as
\begin{equation}
\label{eq:delay}
    \mathbf{D}_{ij} \phi_j(t) = \phi_j(t - d_{ij}(t)), 
\end{equation}
where $d_{ij}(t)$ is the light travel time between the spacecraft. The variability in the travel time is owed to the flexing motion of the constellation around its barycenter. 

Due to data budget limitations the initial sampling rate of \qty{80}{\mega\Hz} must be drastically decimated down to \qty{4}{\Hz}. For efficiency, the entire decimation and filtering operation is done in several steps as shown in \fref{fig:onboard_processing}, each stage consists of an anti-aliasing filter $\filter$ preceding a decimator $\mathbf{S}$. First, a \gls{cic} filter is used to decimate the sampling rate from $\sim\qty{80}{\mega\Hz}$ to \qty{608}{\Hz}. Then, a series of \gls{fir} filters are employed to achieve a final sampling rate of \qty{4}{\Hz}~\cite{Yamamoto:2021ujg}. We indicate measurements that are affected by this decimation stage by a bar
\begin{equation}
    \bar{\eta}_{ij} = \mathbf{S}\mathbf{F} \eta_{ij} \,.
\end{equation}
The design of the decimation stage used in this study is presented in \ref{app:filter_design}. To save computational resources on board the spacecraft we optimize the design by minimizing the number of taps per filter while still retaining sufficient stop-band suppression to limit the effect of aliasing.

The filtering operation can be represented by the convolution of the filter's impulse response $h_\filter(\tau)$ with the data $x(t)$. Formally, we have
\begin{equation}
    \filter x(t) = (h_\filter * x)(t) = \int_\mathbb{R} \! h_\filter(\tau) \cdot x(t - \tau) \,\mathrm{d}\tau . \label{eq:convolution}
\end{equation}
For \gls{fir} filters that operate on time-discrete data sampled with a cadence of $T_\mathrm{s}$ the impulse response and its Fourier transform is given by
\begin{align}
    h_\filter(\tau) &= \sum_m h_m \delta(\tau - m \cdot T_\mathrm{s}) , \label{eq:impulse_response} \\
    \tilde h_\filter(f) &= \sum_m h_m e^{- 2\pi i f m T_\mathrm{s}} , \label{eq:transfer_function}
\end{align}
where $h_m$ are the filter coefficients and $\delta(\tau)$ denotes the Dirac-delta distribution~\cite{Staab:2023qrb}.

Note that we do not account for the decimation operation $\mathbf{S}$ in this study and perform the modeling of the noise residuals and also the numerical experiments in \sref{sec:Results} at the final sampling rate of \qty{4}{\Hz}.
To model the anti-aliasing filters shown in \fref{fig:onboard_processing}, we iteratively build up the coefficients for an equivalent filter that encompasses all four \gls{fir} stages running at the ``high'' sampling frequency of \qty{608}{\Hz}. This is done by upsampling (filling with zeros) the filter coefficient array of the final filter, i.e. \gls{fir} 4, by the decimation factor of the preceding stage, i.e. 2. Then, the upsampled coefficients are convolved with the coefficients of that stage. This procedure is repeated until we reach the first \gls{fir} filter. Finally, to obtain an effective filter that runs at the final rate of \qty{4}{\Hz} we decimate the upsampled coefficient array by the total decimation factor $19 \cdot 2^3$. The in-band response of this effective filter is almost identical to the true transfer function of the overall filter chain. Since the laser noise residuals discussed in this work are only sensitive to the in-band response of the filter chain we can simplify the modeling and simulation by assuming the effective filter running at the final sampling rate of \qty{4}{\Hz}. This is further explored in \ref{app:filter_design} and more thoroughly in~\cite{Staab2023}.

\section{Time Delay Interferometry}
\label{sec:tdi}

\begin{figure}
    \centering
    \includegraphics[width=0.5\linewidth]{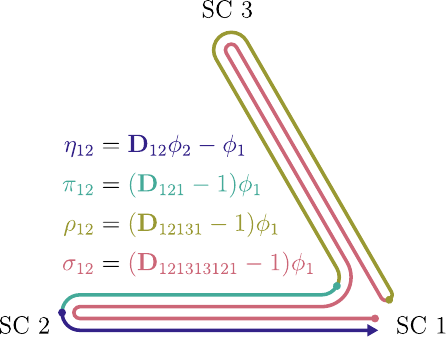}
    \caption{Representation of the construction of the virtual beams between spacecraft required for the second-generation Michelson combination $X_2$. Each color is tracing the ``long arm'' of the intermediary variables defined in \eref{eq:pi} to \eref{eq:sigma}.}
    \label{fig:intervars}
\end{figure}

\Acrlong{tdi} is a post-processing technique to suppress laser noise. It forms virtual near-equal-arm interferometers by linearly combining time-shifted copies of the interferometric measurements; the inter-spacecraft interferometer, reference interferometer and test mass interferometer. 
The TDI algorithm uses estimates of flight-time delays in beam propagation between the spacecraft to time-shift data. The time-delay operation is implemented by interpolating the discrete data sampled onboard and then re-evaluating it on the shifted time grid during post-processing. While this method allows us to apply arbitrary delays on our discrete data, we have to account for noise arising from errors in our chosen interpolator. These  errors have been previously discussed in~\cite{phdthesisOH},~\cite{Staab:2023qrb} and more recently in~\cite{Staab:2024nyo}. Therefore to avoid redundancy, we omit this analysis from our study.

As introduced in \sref{sec:onboard}, we skip the subtraction of spacecraft motion and the reduction to three lasers. Instead, we directly start from the single-link differential phase measurement $\eta_{ij}$, illustrated by the purple path from spacecraft $2$ to spacecraft $1$ in \fref{fig:intervars}. To effectively cancel out the laser noise present in the laser phases $\phi_i$, the six single-link measurements $\eta_{ij}$ are combined to construct virtual beam paths that travel around the constellation and, upon recombination, represent near-equal-arm interferometers. As opposed to ground-based detectors the counter-propagating beams have to take multiple round-trips to compensate unequal inter-spacecraft distances. A commonly used \gls{tdi} variable is the second-generation Michelson combination which constructs a path shown in \fref{fig:intervars}; equivalent to the laser traversing the path to each distant spacecraft twice. It accommodates linear variations in the inter-spacecraft distance of the order of \qty{10}{\m\per\s} and is sufficient for reducing the laser frequency noise below secondary noises for the current design of \gls{lisa}~\cite{Tinto:2003vj,Shaddock:2003dj}.

The construction of the final variable is factorized into intermediary variables for numerical efficiency~\cite{Staab:2023qrb}. Each intermediary variable represents an interferometer itself with a ``short'' and a ``long'' arm. The long arm traces round-trip paths of increasing length as depicted in \fref{fig:intervars}. In the first step we create a two-spacecraft round-trip measurement shown in cyan. It is computed from the single-link measurements $\eta_{ij}$ as
\begin{equation}
    \pi_{ij} = \eta_{ij} + \mathcal{D}_{ij}\eta_{ji} \sim (\delay_{iji} - 1) \phi_i \label{eq:pi},
\end{equation}
where we denote the post-processing delay operator as $\mathcal{D}$. This operation makes use of interpolation to evaluate the discretely sampled data at the delayed time~\cite{Shaddock:2004ua,Staab:2024nyo}. This introduces noise in the system due to interpolation errors~\cite{Staab:2023qrb}. In \eref{eq:pi} we re-expressed the laser phase to leading order in terms of the ``true'' propagation delay.

In \eref{eq:pi} we make use of index contraction for nested delays. For the propagation delay the simple relation $\mathbf{D}_{ijk} = \mathbf{D}_{ij} \mathbf{D}_{jk}$ holds and we derive that the nested time delay is given by
\begin{equation}
    d_{ijk} = d_{ij} + \mathbf{D}_{ij} d_{jk} .
\end{equation}
However, the equivalence atomic and nested delay operators does not hold true for the post-processing delays \textit{i.e.} $\mathcal{D}_{ijk} \neq \mathcal{D}_{ij} \mathcal{D}_{jk}$. The reason for this is that the operations $\mathcal{D}_{ijk}$ and $\mathcal{D}_{ij} \mathcal{D}_{jk}$ produce dissimilar interpolation errors; most notably the second form interpolates twice. The effect of factoring of delay operators is non-trivial and discussed in more detail in~\cite{Staab:2023qrb}. 

To achieve virtual round-trip paths as required for the second-generation Michelson combination $X_2$ we define the $\rho_{ij}$ and $\sigma_{ij}$ variables that are also illustrated in \fref{fig:intervars} in sand and rose, respectively. They are formally given by
\begin{align}
    \rho_{ij} &= \pi_{ij} + \mathcal{D}_{iji} \pi_{ik} \sim (\delay_{ijiki} - 1) \phi_i , \label{eq:rho} \\
    \sigma_{ij} &= \rho_{ij} + \mathcal{D}_{ijiki} \rho_{ik} \sim (\delay_{ijikikiji} - 1) \phi_i , \label{eq:sigma} \\
    X_2 &= \sigma_{13} - \sigma_{12} \sim [[\delay_{131}, \delay_{121}], \delay_{12131}] \phi_1 . \label{eq:X2}
\end{align}
In the last line we define $X_2$ as the difference of the two respective round-trip variables effectively canceling the ``short'' arm and leaving behind a virtual interferometer that interferes two beams that have traveled the paths $1 \to 2 \to 1 \to 3 \to 1 \to 3 \to 1 \to 2 \to 1$ and $1 \to 3 \to 1 \to 2 \to 1 \to 2 \to 1 \to 3 \to 1$. For the sake of brevity we represent the final difference of delay operators as a second order delay commutator. From this we can directly follow that laser frequency noise is canceled up to second order in inter-spacecraft velocities and up to first order in accelerations~\cite{Staab:2023qrb,Bayle:2021mue} as the differential light travel time between the two virtual beams reads
\begin{equation}
    \Delta d_{X_2} = ( d_{131}\,\dot{d}_{121} - d_{121}\,\dot{d}_{131})(\dot{d}_{121} + \dot{d}_{131}) - (d_{131}\ddot{d}_{121} - d_{121}\ddot{d}_{131})(d_{121} + d_{131}). \label{eq:mismatch}
\end{equation}
For Earth trailing orbits~\cite{Martens:2021phh}, this is of the order of \qty{e-12}{\s}~\cite{Hartwig:2022yqw}. The combination $Y_2$ and $Z_2$ that complete the triple can be derived by cyclic permutation of the indices.

Until now, we have neglected the inclusion of the on-board processing of the single-link measurement $\eta_{ij}$ discusses in \sref{sec:onboard} for readability. In reality, the starting point for \gls{tdi} is the filtered and decimated single-link variable, which is represented as
\begin{equation}
\label{eq:eta_decimated}
    \bar{\eta}_{ij} = \mathbf{S}\mathbf{F} \mathbf{D}_{ij}\phi_{j} - \mathbf{S}\mathbf{F} \phi_{i} .
\end{equation}
It follows that formulation of all consequent intermediary variables above remains valid,  with each variable now carrying the bar notation.

\begin{figure}
    \centering
    \includegraphics[width=0.8\textwidth]{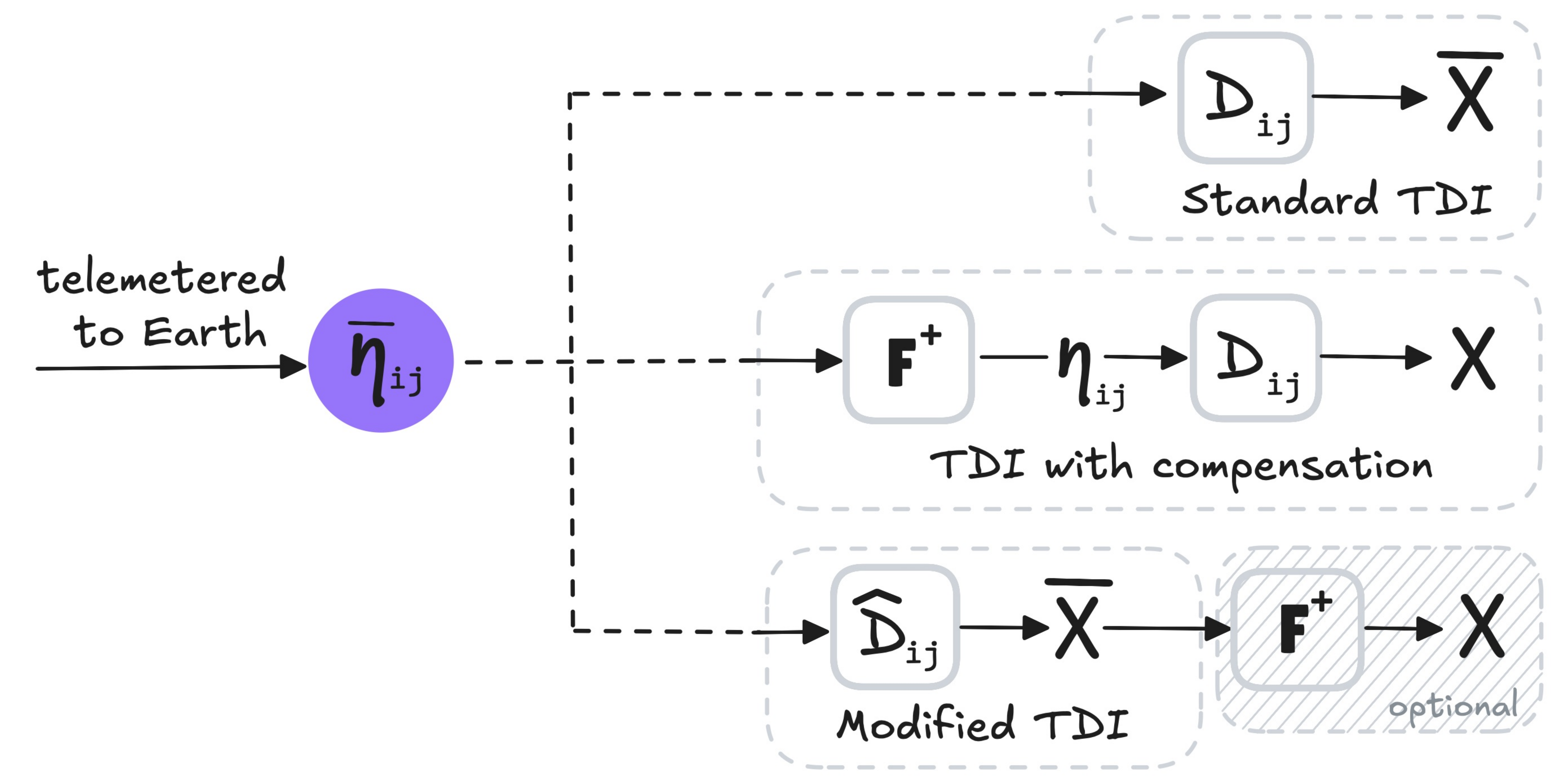}
    \caption{A diagrammatic representation of the different on-ground processing \gls{tdi} topologies; standard \gls{tdi}, \gls{tdi} with compensation and modified \gls{tdi}, starting with the decimated single-link variable $\bar{\eta}_{ij}$.}
    \label{fig:tdi_topologies}
\end{figure}
\subsection{Standard TDI}

Using the decimated single-link variable as an input for \gls{tdi} results in residual laser noise~\cite{Bayle:2018hnm,Staab:2023qrb}. This baseline \gls{tdi} implementation is referred to as standard \gls{tdi} in the subsequent analysis and illustrated in \fref{fig:tdi_topologies}.  The reason for this is the non-commutativity of the filtering and decimation operation with the delay operation. To identify the residuals, we restore the algebraic structure of \eref{eq:eta} by switching around the order of the decimation stage and the delay operation
\begin{align}
    \bar{\eta}_{ij} = \mathbf{D}_{ij}\bar{\phi}_{j} - \bar{\phi}_{i} + \underbrace{[\mathbf{S}\mathbf{F},\mathbf{D}_{ij}]\phi_{j}}_{\text{residual}} ,\label{eq:eta_res}
\end{align}
where $\bar{\phi}_i = \mathbf{S}\mathbf{F} \phi_i$ as indicated above.
The re-arrangement of operators gives rise to a commutator term which is not suppressed in \gls{tdi}. This residual can be split into two contributions

\begin{equation}
\label{eq:commutator}
    [\mathbf{S}\mathbf{F},\mathbf{D}_{ij}] = \underbrace{[\mathbf{S},\mathbf{D}_{ij}]}_{\text{aliasing}}\mathbf{F}+
    \mathbf{S}\underbrace{[\mathbf{F},\mathbf{D}_{ij}]}_{\hspace{-5ex}\text{flexing filtering}\hspace{-5ex}}, 
\end{equation}
where the former residual is attributed to aliasing. This residual can be suppressed by ensuring a sufficiently large attenuation in the stop-band. In the subsequent analysis, the decimation operator $\mathbf{S}$ is excluded from calculations. This is without consequence as the aliasing effect is out-of-scope for our work and the ``aliased'' flexing filtering contribution of the first commutator in \eref{eq:commutator} is negligible. The latter laser noise residual is due to the flexing-filtering effect.

The current baseline design for the \gls{lisa} mission foresees taking the filtered single-link measurements $\bar\eta_{ij}$ directly as an input for \gls{tdi} as defined in \eref{eq:pi} to \eref{eq:X2}. This yields
\begin{equation}
    \bar{X}_2^{\mathbf{D}} \approx [[\mathbf{D}_{131},\mathbf{D}_{121}],\mathbf{D}_{12131}]\bar{\phi}_{1} + \underbrace{\delta\!X_2^{\mathcal{D}}}_{\text{interpolation}} + \underbrace{\delta\!X_2^{[\mathbf{F},\mathbf{D}]}}_{\text{flexing-filtering}} , \label{eq:TDI_standard}
\end{equation}
where the secondary terms are laser noise residuals due to interpolation errors and the flexing-filtering effect (further detailed in \sref{sec:residuals}). These residuals need to be controlled such that they are only a minor contribution in the overall noise budget. To suppress the interpolation residual the post-processing delay operation has to implement a high-accuracy interpolation method. A common choice for this task is Lagrange interpolation as it performs very well at low frequencies~\cite{Shaddock:2004ua}. Recently, a new class of interpolation methods was studied that reduces the computational cost of the operation~\cite{Staab:2024nyo}. To limit the residual due to the flexing-filtering effect, the filter design needs to be sufficiently flat in band. While several configurations have been proposed to meet this ``flatness'' requirement~\cite{Bayle:2019}, the filter design parameters used in this study are presented in \ref{app:filter_design}.

Flat filter designs increase the computational cost of on-board processing. This is undesirable given the limited resources available on the spacecraft. To relax the computational demand of the on-board filter chain, the flatness-requirement on the anti-aliasing filter can be dropped and the in-band response can be corrected on ground. This leaves sufficient stop-band attenuation to limit the aliasing effect as the driving design criterion. We will explore this option in the following section.

\subsection{TDI with Compensation}
\label{sec:TDI_w_comp}

To correct for the non-unity-gain response of the on-board anti-aliasing filter we propose to apply a compensation filter on-ground prior to \gls{tdi}~\cite{Staab2023}, as depicted in \fref{fig:tdi_topologies}. Ideally, the transfer function of this filter is equal to the inverse response of the on-board filter chain $\filter$ to flatten out the response completely. However, in practice, we can only design a pseudo-inverse of the anti-aliasing filter $\filter^+$, as a true inverse would require an infinitely long filter. Furthermore, it is desirable to design a filter that has a minimal amount of coefficients to reduce computational cost and the overall group delay of the filter chain.

After application of the compensation filter we (almost) recover the original single-link measurements before decimation as
\begin{equation}
     \mathbf{F}^+\bar{\eta}_{ij} = \underbrace{\mathbf{F}^+\mathbf{F}}_{\approx \mathds{1}}\,(\mathbf{D}_{ij}\phi_{j}-\phi_{i}) \approx \eta_{ij} .
\end{equation}
Here, $\filter^+ \filter$ is the effective filter that acts on the measurement and that needs to be taken into account when evaluating the flexing-filtering effect. Any remaining non-flatness will amount to a residual flexing-filtering coupling.

The second-generation Michelson variable $X_2$ with compensation can be expressed as
\begin{equation}
    X_2^{\mathbf{D}} \approx [[\mathbf{D}_{131},\mathbf{D}_{121}],\mathbf{D}_{12131}]\phi_{1} + \underbrace{\delta\!X_2^{\mathcal{D}}}_{\text{interpolation}} + \underbrace{\delta\!X_2^{[\filter^+ \filter,\mathbf{D}]}}_{\text{(residual) flexing-filtering}} ,
    \label{eq:TDI_with_comp}
\end{equation}
where we identify laser noise residuals due to interpolation errors and the flexing-filtering effect as in \eref{eq:TDI_standard}. However, the coupling of the flexing-filter effect is reduced as $\filter^+ \filter$ is much more flat by design.

\subsection{Modified TDI}
\label{sec:tdi_wo_comp}

The order of applying the compensation filter and performing \gls{tdi} is not fixed. Analogous to \gls{tdi} without clock synchronization~\cite{Hartwig:2022yqw}, where \gls{tdi} is performed prior to clock synchronization, the order of the compensation filter and \gls{tdi} can be interchanged, as depicted in \fref{fig:tdi_topologies}. Reducing laser noise as the most dominant noise contribution early in the pipeline has the advantage that subsequent processing steps that are sensitive to laser noise cannot couple anymore. This means the requirements on an optional compensation filter post-\gls{tdi} are not driven by the flexing-filtering coupling anymore which possibly reduces the length of the filter drastically. We call this approach ``modified \gls{tdi}'' because the input signal is the same decimated interferometric signal as in standard \gls{tdi}, without the ``removal'' of the anti-aliasing filter using the compensation filter. 

Let us now derive the expression for the modified delay operator in \gls{tdi} that accounts for the anti-aliasing filter. We begin with the definition of the filtered single-link measurement $\bar\eta_{ij}$ defined in \eref{eq:eta_decimated} and insert the unity operation $\mathds{1} = \filter^{-1} \filter$ after the delay operation. We find,
\begin{equation}
    \bar{\eta}_{ij} = \underbrace{\filter\mathbf{D}_{ij}\filter^{-1}}_{\widehat{\mathbf{D}}_{ij}}\filter\phi_{j} - \filter\phi_{i} = \widehat{\mathbf{D}}_{ij}\bar{\phi}_{j} - \bar{\phi}_{i}
    \label{eq:delay_modified}
\end{equation}
where we recover the algebraic structure of \eref{eq:eta} by recasting the delay operator as $\widehat{\mathbf{D}}_{ij} = \filter \mathbf{D}_{ij} \filter^{-1}$ and the laser phase as $\bar{\phi}_i = \filter \phi_i$. Performing \gls{tdi} as given in \eref{eq:pi} to \eref{eq:X2} using the modified delay operator yields
\begin{align}
    \bar X_2^{\widehat{\mathbf{D}}} &\approx [[\widehat{\mathbf{D}}_{131},\widehat{\mathbf{D}}_{121}],\widehat{\mathbf{D}}_{12131}]\bar\phi_{1} + \delta\!X_2^{\widehat{\mathcal{D}}} \\
    &= \filter [[\mathbf{D}_{131},\mathbf{D}_{121}],\mathbf{D}_{12131}] \phi_{1} + \underbrace{\delta\!X_2^{\widehat{\mathcal{D}}}}_{\hspace{-5em}\text{interpolation + correction}\hspace{-5em}} . \label{eq:TDI_wo_comp}
\end{align}
As expected we recover the filtered second-generation Michelson combination alongside laser noise residuals due to mismodeling of the modified delay operator. Here, we differentiate between errors due to interpolation of the discrete time series and approximation errors that concern the correction for the filter. We make this distinction more appreciable in the next section, where we discuss the numerical implementation of the modified delay operation.

\section{The Modified Delay Operator}
\label{sec:derivation}

\begin{figure}
    \centering
    \includegraphics{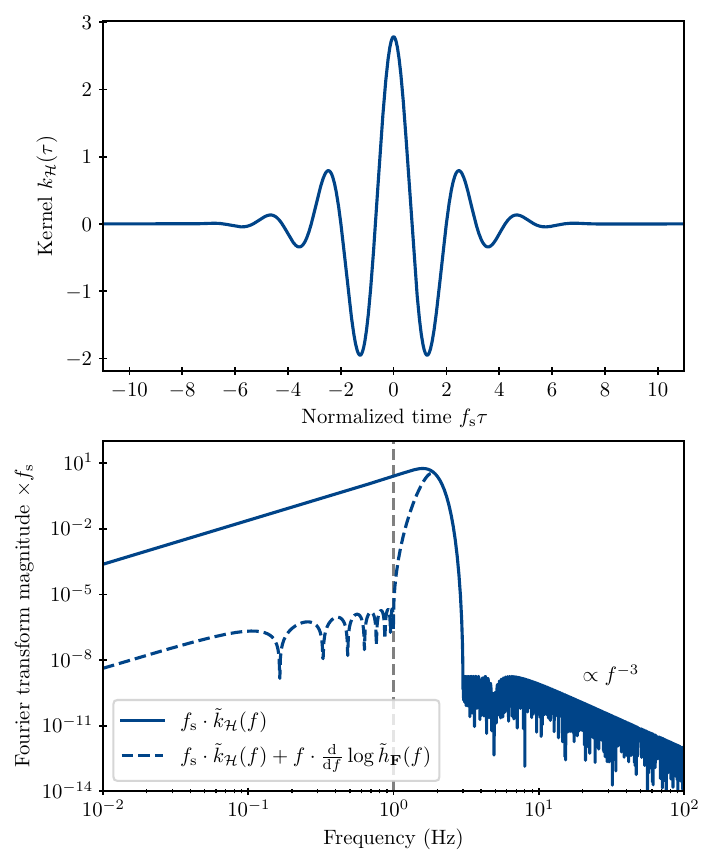}
    \caption{Illustration of the kernel function. The upper plot shows the time-domain representation. The lower plot depicts its normalized Fourier transform (solid) and deviation from the exact response (dashed). In band the Fourier transform follows closely the desired function $-f \cdot \frac{\mathrm{d}}{\mathrm{d}f} \log \tilde{h}_\mathbf{F}(f)$ as required by \eref{eq:worst_case_error}. The dashed gray vertical line marks the upper end of the \gls{lisa} band.}
    \label{fig:kernel}
\end{figure}

In \eref{eq:delay_modified} we have defined the modified delay operator that needs to be considered for the modified \gls{tdi} scheme. To use it in numerical calculations working with discrete-time data, we need to derive an appropriate numerical implementation. Therefore, we adapt the method for implementing the usual post-processing delay operation as a time-varying FIR filter~\cite{Staab:2024nyo} and introduce an appropriate correction term. To identify this correction term we insert the filter-delay-commutator in \eref{eq:delay_modified}. Then, we recognize that for slowly varying delays (small delay derivatives $\dot d$) we can expand the filter-delay-commutator~\cite{Staab:2023qrb} and find
\begin{align}
    \widehat{\mathbf{D}} &= \filter\delay\filter^{-1} ,\\
    &= \delay + [\filter,\delay]\filter^{-1} ,\\
    &\approx \delay + \dot d \cdot \underbrace{\delay \frac{\mathrm{d}}{\mathrm{d}t}\mathbf{G} \filter^{-1}}_{\mathbf{H}} .
    \label{eq:d_hat}
\end{align}
Here, $\mathbf{G}$ is a filter with an impulse response of $h_{\mathbf{G}}(\tau) = \tau \cdot h_\filter(\tau)$. The modified delay operator is a sum of the normal delay operation and a small correction that is proportional to $\dot d(t)$. The effective operator $\mathbf{H}$, referred to as the flexing-filtering correction, acts similar to a delay operator with an altered (non-unity) response in band. Therefore, we define its action on discrete data similarly to the pure delay operation as
\begin{equation}
    y(n T_\mathrm{s}) = \sum_{m=-\infty}^{\infty} x\left((n - m) T_\mathrm{s}\right) \cdot k_\mathcal{H}\!\left(m T_\mathrm{s} - d\right) ,\label{eq:discrete_convolution}
\end{equation}
where $k_\mathcal{H}(\tau)$ is the so-called kernel function; an even function of finite width to allow the sum to run over only a finite number of indices. When applying the full modified delay operator $\widehat{\mathcal{D}}$, the kernel function for the delay operation $k_\mathcal{D}(\tau)$ and one for the correction $k_\mathcal{H}(\tau)$ can be combined for computational efficiency (a single convolution) to form a single effective kernel function
\begin{equation}
    k_{\widehat{\mathcal{D}}}(\tau, t) = k_\mathcal{D}(\tau) + \dot d(t) \cdot k_\mathcal{H}(\tau) ,
\end{equation}
that is explicitly dependent on time as it involves the delay derivative $\dot d(t)$.

In the following we discuss how to design $k_\mathcal{H}(\tau)$ appropriately to approximate the operator $\mathbf{H} \approx \mathcal{H}$. To derive the frequency response of the convolution defined in \eref{eq:discrete_convolution} we analyze the (discrete-time) Fourier transform of the kernel function
\begin{align}
        \tilde h_\mathcal{H}(f;d) &= \sum_m k_\mathcal{H}\left(m T_\mathrm{s} - d\right) \cdot e^{-2\pi i m T_\mathrm{s} f} ,\\
        &= f_\mathrm{s} \sum_m \tilde k_\mathcal{H}\left(f - m f_\mathrm{s}\right) \cdot e^{-2\pi i (f - m f_\mathrm{s}) d} . \label{eq:kernel_response}
\end{align}
The first form can be conveniently used to calculate the response as the sum only runs over a finite number of indices. The second form links the Fourier transform of the kernel function to the response. Ideally, $\tilde h_\mathcal{H}(f)$ follows the exact frequency response of the operator $\mathbf{H}$ which is given by
\begin{equation}
    \tilde h_\mathbf{H}(f;d) = -e^{-2\pi i f d} \cdot f \cdot \frac{\mathrm{d}}{\mathrm{d}f} \log \tilde{h}_\mathbf{F}(f) .\label{eq:desiredcomp}
\end{equation}
This results is computed by multiplication of the individual transfer functions of each operator appearing in the definition of $\mathbf{H}$ in \eref{eq:d_hat}. For reference the individual transfer functions are given by $\mathbf{D} \to e^{-2\pi i f d}$, $\frac{\mathrm{d}}{\mathrm{d}t} \to 2 \pi i f$, $\mathbf{G} \to \frac{i}{2\pi}\frac{\mathrm{d}}{\mathrm{d}f}\tilde h_\filter(f)$ and $\filter^{-1} \to \tilde h_\filter^{-1}(f)$.

The following upper bound on the deviation of the approximate response to the exact one can be derived by using \eref{eq:kernel_response}, \eref{eq:desiredcomp} and the triangle inequality
\begin{equation}
        \left|\tilde h_\mathcal{H}(f;d) - \tilde h_\mathbf{H}(f;d)\right| \le \left|f_\mathrm{s} \tilde k_\mathcal{H}(f) + f \cdot \frac{\mathrm{d}}{\mathrm{d}f} \log \tilde{h}_\mathbf{F}(f)\right| + \sum_{m \neq 0} \left|f_\mathrm{s} \tilde k_\mathcal{H}\left(f - m f_\mathrm{s}\right)\right| .
    \label{eq:worst_case_error}
\end{equation}
From here we can derive conditions for the Fourier transform of the kernel function $\tilde k_\mathcal{H}(f)$. First, it should follow $- f \cdot \frac{\mathrm{d}}{\mathrm{d}f} \log \tilde{h}_\mathbf{F}(f)$ in band. Second, it should vanish outside of the band to suppress ``aliased'' contributions originating from the sum on the right-hand side of the equation.

The design procedure for a suitable kernel function $\tilde k_\mathcal{H}(f)$ can be adapted from methods recently developed in~\cite{Staab:2024nyo} for interpolation. Likewise, we base the functional form of $\tilde k_\mathcal{H}(f)$ on the newly introduced class of highly versatile kernel functions called cosine-sum kernels. They are defined as
\begin{equation}
    k_\mathcal{H}(t) = \rect\left(\frac{t}{NT_\mathrm{s}}\right)\sum_{n=0}^{N-1} a_n \cdot \cos\left(2\pi f_\mathrm{s} \frac{n}{N}t\right) . \label{eq:cosine-sum}
\end{equation}
Here, $N$ is a even integer that specifies the width of the kernel and $a_n$ denote an array of $N$ coefficients that need to be optimized to obtain the desired response. We choose a kernel width of $N=22$ (which is equivalent to the number of taps in the filter applied in \eref{eq:discrete_convolution}) to match that of the optimized interpolation kernel recently published in~\cite{Staab:2024nyo}.

As derived above we require that the Fourier transform of the kernel function $\tilde k_\mathcal{H}(f)$ follows the true response $\tilde h_\mathbf{H}(f)$ (excluding the complex exponential) at in-band frequencies and vanishes in the stop band to limit the effect of aliasing. Therefore, we define the desired response as
\begin{equation}
    D(f) =
    \begin{cases*}
        -f \cdot \frac{\mathrm{d}}{\mathrm{d}f} \log \tilde{h}_\mathbf{F}(f) & if $\qty{0}{\Hz} \le f \le \qty{1}{\Hz}$, \\
        0 & if $\qty{3}{\Hz} \le f \le \infty$.
    \end{cases*}
\end{equation}
We do not put additional constraints on frequencies between \qty{1}{\Hz} and \qty{3}{\Hz} as they are either out-of-band or fold to out-of-band frequencies.

Analogously to~\cite{Staab:2024nyo}, we set up the weighted error function in the frequency domain as
\begin{equation}
    E(f) = W(f) \left(\tilde k_\mathcal{H}(f) - D(f)\right) ,
\end{equation}
where $W(f)$ denotes frequency-dependent weights. The maximum error over all frequencies of the considered domain $f \in [0, \qty{1}{\Hz}] \cup [\qty{3}{\Hz}, \infty)$ must be minimized. After refactoring and rearranging the equation we identify a weighted Chebyshev approximation problem that can be solved using a slightly modified version of the Parks-McClellan algorithm. The algorithmic details of the optimization are discussed in \ref{app:correction_design}.

The resulting kernel function is displayed in \fref{fig:kernel}. The upper plot shows the time-domain representation $k_\mathcal{H}(\tau)$ which extends (in normalizes time) over \num{22} samples. As required, it is exhibits even symmetry and falls off to zero at the boundaries. The lower plot depicts the magnitude of the Fourier transform (normalizes by the factor $f_\mathrm{s}$) and its deviation from the exact response in dashed. In band ($\qty{0}{\Hz} < f < \qty{1}{\Hz}$) the Fourier transform follows closely the exact response. At stop-band frequencies ($\qty{3}{\Hz} < f$) it is well suppressed to limit aliasing of those components.

\section{Laser Noise Residuals}
\label{sec:residuals}

Each of the three topologies discussed in \sref{sec:tdi} introduces residual laser noise into the final \gls{tdi} variable. Many of those have already been identified in previous literature~\cite{Staab:2023qrb}; flexing-filtering coupling~\cite{Bayle:2018hnm}, the aliasing effect, coupling of ranging errors and the effect of interpolation errors. Here, we focus on the residuals that involve the anti-aliasing filter $\filter$. Hence, we neglect the aliasing effect and assume exact ranges.

Using the assumptions stated above, standard \gls{tdi} as derived in \eref{eq:TDI_standard} is limited by interpolation errors $\delta\!X_2^{\mathcal{D}}$ and the flexing-filtering coupling $\delta\!X_2^{[\mathbf{F},\mathbf{D}]}$. Interpolation errors arise as the post-processing delay operator deviates from the true delay operation $\diff = \fdelay - \delay$. The resulting residuals formally read
\begin{align}
    \delta\!X_2^{\mathcal{D}} &= \tdi \Big(\big(\delay (\diff_{13} - \diff_{12}) + (\diff_{121} - \diff_{131})\big) \bar\phi_1 + \diff_{12} \bar\phi_2 - \diff_{13} \bar\phi_3 \Big) , \label{eq:interpolation_error}\\
    \delta\!X_2^{[\mathbf{F},\mathbf{D}]} &= \tdi \Big( \delay \big([\filter, \delay_{31}] - [\filter, \delay_{21}]\big) \phi_1 - [\filter, \delay_{12}] \phi_2 + [\filter, \delay_{13}] \phi_3 \Big) . \label{eq:flexing_filtering}
\end{align}
Here, we assume equal arms to simplify the models similar to \cite{Staab:2023qrb}. The operator $\tdi$ denotes a common factor in front \gls{tdi} residuals. For the second-generation Michelson variable it is defined as
\begin{align}
    \tdi &= (1 - \delay^4)(1 - \delay^2), \\
    |\tilde{\mathbf{C}}| &= 4 \sin(2\pi f d_0) \sin(4\pi f d_0) .
\end{align}
In the second line we quote the magnitude of its transfer function, commonly know as the ``\gls{tdi} transfer function''~\cite{Staab:2023qrb}.
To isolate the residuals related to the anti-aliasing filter in our study we minimize the noise contribution from interpolation error by choosing the more expensive Lagrange interpolation scheme with $N=62$ coefficients.

Therefore, the dominant residual in standard \gls{tdi} becomes the flexing-filtering coupling. As derived in \eref{eq:eta_res} and \eref{eq:commutator}, the coupling is due to the non-commutativity of the filtering and delay operation. As shown in~\cite{Bayle:2018hnm} and~\cite{Staab:2023qrb}, the commutator can be expanded up to first order in the delay derivatives. The \gls{asd} of \eref{eq:flexing_filtering} is therefore given by
\begin{equation}
    \sqrt{S_{\delta\!X_2}^{[\mathbf{F}, \mathbf{D}]}(f)} = 2 \big|\tilde\tdi\big| \left| \frac{\bar{\dot{d}}}{2\pi} \cdot \frac{\mathrm{d}\tilde{h}_{\mathbf{F}}(f)}{\mathrm{d}f} \right| \sqrt{S_{\dot \phi}(f)} .
    \label{eq:flexing_filtering_asd}
\end{equation}
The flexing-filtering residual scales linearly with the effective delay derivative $\bar{\dot d}$ and the derivative of the transfer function of the anti-aliasing filter, i.e. the non-flatness of the FIR filter in the pass band. The definition of $\bar{\dot d}$\footnote{Here, the bar denotes averaging and not decimation.} slightly differs from \cite{Staab:2023qrb} due to the three-laser configuration which we assume in this article. Here, it is defined as
\begin{equation}
    \bar{\dot d} = \sqrt{\frac{\dot d_{12}^2 + \dot d_{13}^2 + (\dot d_{21} - \dot d_{31})^2}{4}} . \label{eq:ddot_effective}
\end{equation}

The flexing-filtering residual puts strict requirements on the on-board filter to have a near unity frequency response.
To limit the computational cost and the group delay of the decimation chain we introduce a so-called compensation filter $\filter^+$ that flattens out the response sufficiently. This compensation filter has \num{7} coefficients and runs at the final rate of \qty{4}{\Hz} and can therefore also be integrated in the on-ground data processing pipeline. The filter design is described in detail in \ref{app:filter_design}. Performing \gls{tdi} with compensation reduces the flexing-filtering effect. The left-over non-flatness ($\mathbf{F}^+\mathbf{F} \neq 1$) of the filter chain still amounts to a ``residual'' flexing-filtering coupling as given in \eref{eq:TDI_with_comp}. The derivation of the \gls{asd} is trivial since we only need to include the transfer function of the compensation filter in \eref{eq:flexing_filtering_asd} by replacing $\tilde h_\filter(f) \to \tilde h_{\filter^+}(f) \cdot \tilde h_\filter(f)$.

Modified \gls{tdi} (see \sref{sec:tdi_wo_comp}) is an alternative approach to mitigate the flexing-filtering effect. Here, the delay operation $\mathcal{D}$ in \gls{tdi} is replaced by the modified delay $\widehat{\mathcal{D}}$ (described in \sref{sec:derivation}) to include the effect of the filter $\filter$. As a result the traditional flexing-filtering effect is averted and only residuals related to interpolation and mismodeling of the correction term $\mathbf{H}$ remain (see \ref{eq:TDI_wo_comp}). The numerical implementation of the modified delay as described in \sref{sec:derivation} is imperfect, giving rise to laser noise residuals. Similar to the interpolation error in \eref{eq:interpolation_error} laser noise couples through the difference between the post-processing and exact modified delay 
\begin{equation}
    \diff = \widehat{\fdelay} - \widehat{\delay} = \underbrace{(\fdelay - \delay)}_\mathrm{interpolation} + \underbrace{\dot d \cdot (\mathcal{H} - \mathbf{H})}_\mathrm{correction} .
\end{equation}
Here, we have expanded the modified delay operators using \eref{eq:d_hat} to identify laser noise residuals due to interpolation errors and due to the mismodeling of the flexing-filtering correction $\mathbf{H}$.

We estimate the \gls{asd} of the correction residual by assuming that the delays are only slowly changing such that we can assume the operations $\mathbf{H}$ and $\mathcal{H}$ to be time-invariant. Then, we can use \eref{eq:worst_case_error} to estimate the worst-case \gls{asd} of the correction residual $\delta\!X_2^{\mathcal{H}}$ over all possible values of the delays. It reads
\begin{equation}
    \begin{split}
        \sqrt{S_{\delta\!X_2}^{\mathcal{H}}(f)} \le 2\big|\tilde\tdi\big| |\dot d| &\left(\left|f_\mathrm{s} \tilde k_\mathcal{H}(f) + f \cdot \frac{\mathrm{d}}{\mathrm{d}f} \log \tilde{h}_\mathbf{F}(f)\right| + \sum_{m \neq 0} \left|f_\mathrm{s} \tilde k_\mathcal{H}\left(f - m f_\mathrm{s}\right)\right|\right) \\ &\cdot |\tilde h_\filter(f)| \sqrt{S_{\phi}(f)} ,
    \end{split}
    \label{eq:correction_asd}
\end{equation}
where $|\dot d|$ denotes a specific combination of the magnitudes of individual delay derivatives given by 
\begin{equation}
    |\dot d| = \frac{|\dot d_{12}| + |\dot d_{13}|}{2} .
\end{equation}

Similar to the flexing-filtering effect, the \gls{asd} of the correction residual scales with the delay derivatives. Also note that this residual couples to laser phase noise, defined as $S_\phi(f) = S_{\dot \phi}(f) / (2 \pi f)^2$.

\section{Simulation Results}
\label{sec:Results}

In this section, we validate the analytical findings of the previous section by running numerical simulations. We generate the interferometric signals using the LISA Instrument~\cite{Bayle:2022okx,bayle_2024_13809621} simulator and compare the residual noise in the second-generation Michelson variable for the three \gls{tdi} topologies given in \sref{sec:tdi} against their analytical models. We assume the three-laser configuration where each spacecraft houses only a single laser source that is locked to a cavity. As a result the beatnote frequency of the inter-spacecraft interferometer can be readily used as an input to \gls{tdi} as it effectively yields the intermediary variable $\eta_{ij}$. Furthermore, we assume the three lasers to be independent with a flat \gls{asd} of
\begin{equation}
    \sqrt{S_{\dot \phi}} = \qty{30}{\Hz\per\sqrt{\Hz}} .
\end{equation}
All other instrument noises are disabled to isolate the laser noise residuals post \gls{tdi}.

The simulation of all measurements is performed at \qty{4}{\Hz} for \qty{25000}{\s}. Usually, the LISA Instrument simulator runs at a higher rate to simulate the physics of \gls{lisa} with greater precision and also include the effect of decimation. However, the residuals we are discussing in this paper are only related to the in-band transfer function of the on-board filter. Therefore, we only apply a single effective ``decimated'' filter (as described in \sref{sec:tdi}) that emulates the full on-board processing stage without actually decimating the sampling rate. To account for the reduced Nyquist frequency when propagating beams between spacecraft we increase the interpolation order (Lagrange) to \num{61}.

To obtain realistic \gls{lisa} dynamics we choose numerical orbits files provided by \gls{esa}~\cite{Martens:2021phh,bayle_2022_7700361}. Furthermore, we set the starting time of our simulation to $t_0 = \qty{2.0813e9}{\s}$ to yield a large effective delay derivative $\bar{\dot d} \approx \qty{2e-8}{\s\per\s}$ and $|\dot d| \approx \qty{2.3e-8}{\s\per\s}$. The LISA Instrument simulator works in units of frequency. Therefore, the time-dependent delay operators used in \gls{tdi} (as derived in \sref{sec:tdi}) have to be updated to account for Doppler shifts. The pure delay operation $\mathbf{D}$ is easily adjusted. According to~\cite{Bayle:2021mue} in addition to the time-delay frequency data must also be multiplied by a Doppler factor. We define the Doppler-corrected delay operator applied to an arbitrary frequency $\nu$ as
\begin{equation}
 \dot{\mathbf{D}} \nu \equiv \left(1 - \dot{d}(t)\right) \cdot \mathbf{D} \nu . \label{eq:doppler_delay}
\end{equation}

The derivation for the modified delay operator is more cumbersome. First, we recognize that in our formalism introduced in \eref{eq:eta_decimated} we can simply replace $\filter \to \filter \frac{\mathrm{d}}{\mathrm{d}t}$ to effectively yield decimated beatnote frequencies. In the next step, we derive the modified delay operator by performing the substitution in \eref{eq:d_hat}. Formally, we have
\begin{align}
    \dot{\widehat{\mathbf{D}}} &= \mathbf{D} + \dot{d}\,\mathbf{D}\frac{\mathrm{d}}{\mathrm{d}t} \left(\frac{\mathrm{d}}{\mathrm{d}t}\mathbf{G} - \filter\right) \left(\filter \frac{\mathrm{d}}{\mathrm{d}t}\right)^{-1} \nonumber \\ 
    &= (1 - \dot{d})\,\mathbf{D} + \dot{d}\,\underbrace{\mathbf{D}\frac{\mathrm{d}}{\mathrm{d}t}\mathbf{G} \filter^{-1}}_\mathbf{H} , \label{eq:doppler_delay_hat}
\end{align}
where we have used the definition of the convolution integral in \eref{eq:convolution} to derive that $\mathbf{G} \to \frac{\mathrm{d}}{\mathrm{d}t} \mathbf{G} - \mathbf{F}$\footnote{As $h_\filter(\tau) \to \dot h_\filter(\tau)$ the impulse response of the $\mathbf{G}$ operator becomes $h_\mathbf{G}(\tau) \to \tau \cdot \dot h_\filter(\tau)$. Then, we express $\tau \cdot \dot h_\filter(\tau) = \frac{\mathrm{d}}{\mathrm{d}\tau} (\tau \cdot h_\filter(\tau)) - h_\filter(\tau)$ in the convolution integral, which concludes the derivation.}. We recover the usual Doppler-corrected delay operator and the identical correction term as for phase units.

\begin{figure}
    \centering
    \includegraphics{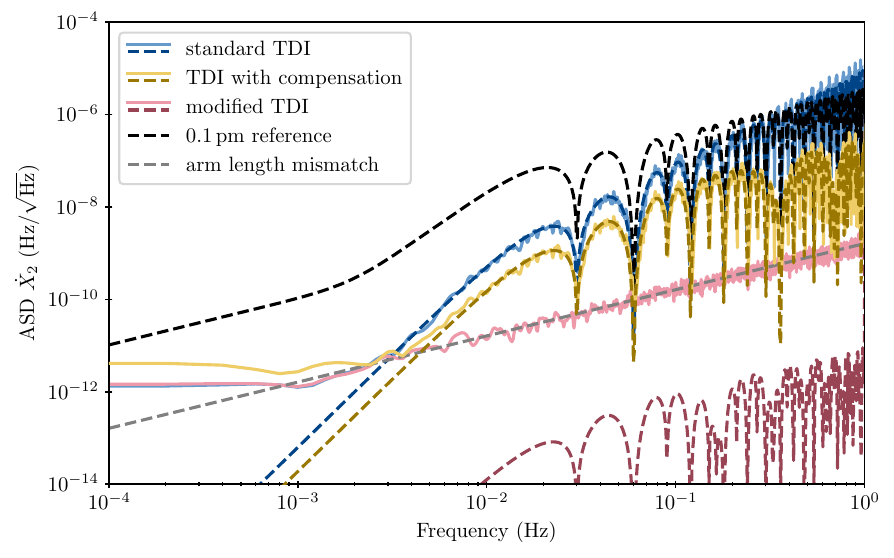}
    \caption{Normalized laser noise residuals of the second-generation Michelson variable $\dot{X}_2$ calculated using standard \gls{tdi} (blue), \gls{tdi} with compensation (yellow) and modified \gls{tdi} (red). Solid lines represent numerical simulations; corresponding models are plotted with dashed lines: flexing-filtering coupling (dashed blue), residual flexing-filtering coupling (dashed yellow), correction residual noise (dashed red) and the fundamental laser noise limit due to arm length mismatch (dashed gray). The \qty{0.1}{\pico\m} reference curve (dashed black) serves as a reference. To make the comparison straight-forward we normalize each \gls{asd} by the transfer function of the corresponding overall effective filter.}
    \label{fig:tdi_performance}
\end{figure}

We use the delay operators defined in \eref{eq:doppler_delay} and \eref{eq:doppler_delay_hat} to calculate the second-generation Michelson variable for the three topologies described in \sref{sec:tdi}. For the numerical implementation the delay operator $\delay$ and flexing-filtering correction $\mathbf{H}$ are replaced by their post-processing equivalents $\mathcal{D}$ and $\mathcal{H}$, respectively. As argued before we use high-order Lagrange interpolation ($N = 62$) to suppress interpolation errors below the noise residuals under study.

In \fref{fig:tdi_performance}, we compare the performance of standard \gls{tdi} (solid blue), \gls{tdi} with compensation (solid yellow) and modified \gls{tdi} (light solid red). These are compared against the \qty{0.1}{\pico\m} reference curve (dashed black), defined as
\begin{align}
    \sqrt{S_{\dot{X}_2}^\mathrm{ref}(f)} = 2 |\tilde{\mathbf{C}}| |2\pi f| \frac{\qty{0.1}{\pico\m\per\sqrt{\Hz}}}{\qty{1064}{\nano\m}} \sqrt{1 + \left(\frac{\qty{2}{\milli\Hz}}{f}\right)^4} .
    \label{eq:noisebudget}
\end{align}
Each of the \gls{asd} models derived in \sref{sec:residuals} are multiplied by the factor $|2 \pi f|$ convert from phase to frequency units. For a straight-forward comparison we normalize each \gls{asd} by the corresponding effective filter transfer function of each \gls{tdi} topology.
 
We observe that the residual for standard \gls{tdi} is well explained by the noise model of the flexing filtering effect (cf. \eref{eq:flexing_filtering_asd}) in dashed blue. Towards higher frequencies, this curve exceeds the \qty{0.1}{\pico\m} reference curve for \gls{lisa} (dashed black). In comparison, the residual for \gls{tdi} with compensation remains below the reference curve, closely described by the predicted residual flexing filtering noise in dashed yellow. The best noise performance is achieved by modified \gls{tdi}. The correction residual noise model (cf. \eref{eq:correction_asd}) is plotted as the red dashed curve. However, this model is drastically underestimating the \gls{psd} of the numerical simulation (light solid red). Therefore, we also plot the fundamental laser noise limit in dashed gray which explains the discrepancy. Its model is given by
\begin{equation}
    \sqrt{S_{\dot{X}_2}(f)} = \Delta d_{X_2} \,|2\pi f| \, |\tilde{h}_\filter(f)| \,\sqrt{S_{\dot{\phi}}(f)} ,
    \label{eq:TDIresidualfreq}
\end{equation}
where $\Delta d_{X_2}$ denotes the arm length mismatch of the second-generation Michelson defined in \eref{eq:mismatch} which in our simulation has an average value of \qty{-8.59e-12}{\s}. In the plot, the factor $|\tilde{h}_\filter(f)|$ does not appear due to the normalization discussed above.

\section{Conclusions} 
\label{sec:conclusions}

In this paper we present a novel approach to correct for the flexing-filtering effect during \gls{tdi}. To test this method, we design a new on-board decimation chain that reduces the computation cost and is compatible with the updated timing architecture of \gls{lisa}. We set sufficient out-of-band attenuation of the anti-aliasing filters as the only design criterion. We show that standard \gls{tdi} which directly uses the decimated interferometric measurement produces a flexing-filtering residual that violates the \qty{0.1}{\pico\meter} reference curve. Therefore, we compare two methods to mitigate the flexing-filtering effect; \gls{tdi} with compensation that was already discussed in the literature and our novel approach dubbed ``modified \gls{tdi}''.

In the \gls{tdi} with compensation topology, a compensation filter $\mathbf{F}^+$ is applied to the decimated data prior to \gls{tdi}. This filter approximates the inverse of the overall transfer function of the decimation chain $\mathbf{F}$ and therefore lifts up the non-unity frequency response of the latter, ensuring a unity gain in-band. Therefore, it successfully reduces the flexing-filtering contribution to fit within the prescribed noise budget. This filter needs a minimum of seven coefficients and can be operated on ground allowing for more relaxed computational requirements. The primary drawback of this method is that the addition of another filter further increases the group delay of the decimation chain. Additionally, the method's efficacy depends on an accurate implementation of the compensation filter as a finite length pseudo-inverse; which amounts to residual flexing-filtering coupling.

As an alternative flexing-filtering mitigation strategy, we propose modified \gls{tdi} which uses a modified \gls{tdi} delay operator. The modification is approximated by adding a small correction scaling with the time-varying delay derivative $\dot{d}$ to the pure delay operator. Similar to the delay operation, the correction is implemented as a time-varying \gls{fir} filter with coefficients drawn from a continuous-time kernel function dependent on the specific design of the on-board anti-aliasing filter. Modified \gls{tdi} has two main advantages over using a compensation filter. First, it requires no additional filters. Second, it has the same ``footprint'' as standard \gls{tdi} since its kernel function has identical width compared to that of the pure delay operation. This results in equivalent data loss around gaps and at the boundary of the the time series. We find that modified \gls{tdi} suppresses flexing-filtering noise by four additional orders of magnitude compared to \gls{tdi} with compensation. 

This result is verified by running numerical simulations and studying the performance of the three \gls{tdi} topologies for the second-generation \gls{tdi} variable $X_2$. We find that the residuals for standard \gls{tdi} and \gls{tdi} with compensation are well explained by our analytical models. For modified \gls{tdi}, the residual is low enough to be swamped by the fundamental laser noise residual due to the flexing of the constellation (arm length mismatch).
The simulations also verify that \gls{tdi} with compensation and modified \gls{tdi} are able to remedy the problem of significant flexing-filtering coupling. 

To verify the efficiency of the approaches, simplifying assumptions should be relaxed. This includes more realistic laser locking schemes instead of the one laser per spacecraft simplification, along with the addition of other primary noise sources in \gls{tdi}. The reduction of clock noise is especially relevant in this context as they rely on auxiliary interferometric measurements that are also affected by the onboard processing chain. Furthermore, the full decimation chain consisting of several stages running at different rates has to be studied and verified in tabletop experiments. Finally, the developed methods need to implemented into existing \gls{lisa} simulation code and prototypes for on-ground processing pipeline.

\ack
The authors thank the LISA Simulation Working Group and the LISA Simulation Expert Group for the lively discussions on all simulation-related activities. In particular, the authors appreciate the timely inputs and support of Aur\'elien Hees and Jean-Baptiste Bayle. We thank Pascal Grafe and  Christoph Bode for their insights on filter development for the \gls{lisa} phasemeter. M.S. gratefully acknowledges support by the Centre National d’\'Etudes Spatiales (CNES). M.S. is supported by the research program of the Netherlands Organisation for Scientific Research (NWO).


\appendix

\section{Anti-Aliasing Filter Design}
\label{app:filter_design}
The anti-aliasing filter $\mathbf{F}$ used in this study is derived from an ad hoc designs of the \gls{lisa} decimation stage. It aims to minimize the usage of on-board computational resources and the total group delay of the entire chain. To achieve this it uses 5 stages of decimation as depicted in \fref{fig:onboard_processing}; a single third-order \gls{cic} filter and four \gls{fir} filters. The \gls{cic} filter is responsible for the initial drastic reduction of the sampling rate from $19 \times 2^{22} \unit{\Hz}$ ($\approx \qty{80}{\mega\Hz}$) to  \qty{608}{\Hz} which corresponds to a decimation factor of $2^{17}$. To reach the final sampling rate of \qty{4}{\Hz} a chain of four successive decimation stages that use \gls{fir} filters is put in place. The first stage decimates by a factor of 19 and the remaining ones by a factor of 2 each\footnote{We choose to split up the remaining factor \num{152} into its prime factors to ease the challenge of the numerical design procedure.}. For an overview see \tref{tab:filter_parameters}.

\begin{table}
    \centering
    \begin{tabular}{l c c c}
        \toprule
        Stage & Decimation factor & \# coefficients & Sampling rates \\
        \midrule
        FIR 1 & 19 & 91 & \qty{608}{\Hz} / \qty{32}{\Hz} \\
        FIR 2 & 2 & 7 & \qty{32}{\Hz} / \qty{16}{\Hz} \\
        FIR 3 & 2 & 9 & \qty{16}{\Hz} / \qty{8}{\Hz} \\
        FIR 4 & 2 & 13 & \qty{8}{\Hz} / \qty{4}{\Hz} \\
        FIR C & 1 & 7 & \qty{4}{\Hz} / \qty{4}{\Hz} \\
        \bottomrule
    \end{tabular}
    \caption{Overview of the \gls{fir} decimation stage properties. The sampling rates correspond to before ($f_\mathrm{s}$) and after ($f_\mathrm{s}'$) decimation is applied.}
    \label{tab:filter_parameters}
\end{table}

The purpose of each decimation stage (anti-aliasing filter + decimator) is to reduce the sampling rate of the data on board while limiting the amount of aliasing of high-frequency laser noise into the \gls{lisa} band. The effect of aliasing is described in~\cite{Staab:2023qrb} as the folding of laser noise power $S(f)$ to in-band frequencies
\begin{equation}
    \rect\left(\frac{f}{f_s'}\right)\sum_{n=0}^{R-1} S^{(n)}(f) ,
    \label{eq:sampling_psd}
\end{equation}
where the so-called aliases $S^{(n)}(f)$ are defined as
\begin{equation}
    S^{(n)}(f) =
    \begin{cases*}
        S(n f_\mathrm{s}' / 2 + f) & if $n$ is even,\\
        S((n + 1) f_\mathrm{s}' / 2 - f) & if $n$ is odd.
    \end{cases*}
    \label{eq:aliases}
\end{equation}
Here, $R$ denotes the decimation factor and $f_\mathrm{s}' = f_\mathrm{s} / R$ the sampling rate after decimation. The $\rect$-function assures that the decimated \gls{psd} is band-limited again.

Similar to \eref{eq:commutator} let us now describe the coupling of the flexing-filtering effect and aliased noise for a chain of multiple decimation stages. For the sake of brevity, we present the derivation for a chain of two filters. The result can easily be generalized. The residual appearing in the $\eta$ variables has the form of a commutator that we can expand into the individual contributions of the two stages by making use of basic commutator rules 
\begin{equation}
    \text{``}[\decimate \filter, \delay]\text{''} =
        (\decimate_2 \filter_2)[\decimate_1 \filter_1, \delay]
        + [\decimate_2 \filter_2, \delay](\decimate_1 \filter_1) .
\end{equation}
Here, ``$[\decimate\filter, \delay]$'' is a hand-wavy notation for the commutator of the full decimation stage and the delay operation. In the next stage we further expand the remaining commutators to distinguish between contributions from the filter-delay commutator (i.e. the flexing-filtering effect) and the decimation-delay commutator (i.e. the aliasing effect). Moreover, we neglect second-order effects, e.g., the coupling of commutators to aliases noise or the aliasing of commutators. As a result, the total filter-delay and sampling-delay commutators read
\begin{align}
    \text{``}\decimate[\filter, \delay]\text{''} &\approx
        \filter_2[\filter_1, \delay] + [\filter_2, \delay]\filter_1 , \\
    \text{``}[\decimate, \delay] \filter\text{''} &\approx
        \filter_2[\decimate_1, \delay]\filter_1
        + [\decimate_2, \delay]\filter_2\filter_1 .
\end{align}
Using the results derived in~\cite{Staab:2023qrb} we write down their respective \glspl{psd}. The \gls{psd} of the overall flexing-filtering effect is given by
\begin{align}
    S_{\delta\nu}^{\decimate[\filter, \delay]}(f) &= \left|\tilde\filter_2 \tilde\dfilter_1 + \tilde\dfilter_2 \tilde\filter_1\right|^2 \cdot \dot d^2 \cdot (2\pi f)^2 \cdot S_{\dot \phi}(f) , \\
    &= \left|\frac{1}{2\pi}\frac{\mathrm{d}}{\mathrm{d}f} \tilde\filter_2 \tilde\filter_1\right|^2 \cdot \dot d^2 \cdot (2\pi f)^2 \cdot S_{\dot \phi}(f) , \label{eq:psd_flexing-filtering_chain}
\end{align}
where $\tilde \filter_i = \tilde h_{\filter_i}(f)$ and $\tilde \dfilter_i = \tilde h_{\filter_i}'(f) / (2 \pi)$ are short-hand notations for the transfer function and its (scaled) derivative, respectively. In the second line we factor out the derivative and recover the intuitive result that the derivative of the total transfer function couples effectively.

For the aliasing effect we proceed analogously. The total aliased laser noise power equals to the sum of the individual contributions and thus reads
\begin{align}
    S_{\delta\nu}^{[\decimate, \delay]\filter}(f) &= 4 |\tilde\filter_2|^2 \sum_{n=1}^{R_1 - 1} (|\tilde \filter_1|^2 S_{\dot \phi})^{(n)}(f) + 4 \sum_{n=1}^{R_2 - 1} (|\tilde\filter_2|^2 |\tilde \filter_1|^2 S_{\dot \phi})^{(n)}(f) \label{eq:psd_aliasing_individual}\\
    &= 4 \sum_{n=1}^{R_1 \cdot R_2 - 1} (|\tilde\filter_2|^2 |\tilde \filter_1|^2 S_{\dot \phi})^{(n)}(f) . \label{eq:psd_aliasing_chain}
\end{align}
Again, the last line collects all individual terms into a single expression. It represents a single decimation operation by the total decimation factor $R_1 \cdot R_2$ of laser noise filtered by the effective total filter.

\begin{figure}
    \centering
    \includegraphics{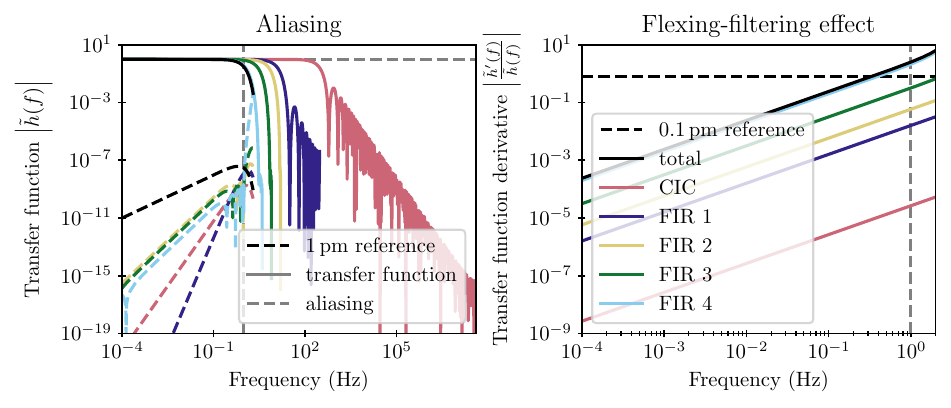}
    \caption{Frequency response of the individual decimation stages including the \gls{cic} stage (rose), and the four successive \gls{fir} stages (indigo, sand, green and cyan). The left plot shows the transfer function of the individual stages as solid lines and aliased contributions as dashed lines (to be scaled by white noise level). The dashed black lines represents the \qty{1}{\pico\meter} reference curve converted to appropriate units assuming a laser noise \gls{asd} of \qty{30}{\Hz\per\sqrt{\Hz}}. The right plot shows the individual and total (solid black) frequency derivative of the transfer function. For comparison the \qty{0.1}{\pico\meter} reference curve is plotted in dashed black. The vertical dashed grey line marks the upper end of the \gls{lisa} band, i.e., \qty{1}{\Hz}.}
    \label{fig:design_cheap_filter}
\end{figure}

Using these results we can formulate requirements on the derivative of the total filter transfer function and the minimum attenuation out of band assuming a worst-case delay derivative of $\dot d = \num{2.5e-8}$ and laser frequency noise with an \gls{asd} of $\sqrt{S_{\dot \phi}} = \qty{30}{\Hz\per\sqrt{\Hz}}$. We require that \gls{psd} estimates in \eref{eq:psd_flexing-filtering_chain} and \eref{eq:psd_aliasing_chain} must be lower than an equivalent displacement noise of \qty{0.1}{\pico\m\per\sqrt{\Hz}} and \qty{1}{\pico\m\per\sqrt{\Hz}}, respectively. To convert to frequency and account for the filter the levels have to be multiplied by the factor $\frac{2 \pi}{\lambda}$ and the filter's transfer function. We find the following conditions,
\begin{align}
    \left|\frac{\tilde h'_\filter(f)}{\tilde h_\filter(f)}\right| &< \frac{(2\pi) \cdot \qty{0.1}{\pico\m\per\sqrt{\Hz}}}{\dot d \cdot \sqrt{S_{\dot \phi}(f)}} , \label{eq:aliasing_requirement}\\
    \sqrt{\sum_{n=1}^{R - 1} (|\tilde h_\filter(f)|^2)^{(n)}(f)} &< \frac{(2 \pi f) \cdot |\tilde h_\filter(f)| \cdot \qty{1}{\pico\m\per\sqrt{\Hz}}}{2 \cdot \sqrt{S_{\dot \phi}(f)}} \label{eq:flexing-filtering_requirement}.
\end{align}
To validate the performance of the filter we plot the right-hand sides of the inequalities above in \fref{fig:design_cheap_filter} as reference curves (black dashed). For a sufficient design the individual contributions have to stay below those curves.

The first decimation stage FIR 1 is responsible for a decimation factor of $R_1 = 19$. To ensure that the filter transfer function has nulls at multiples of the sampling rate after decimation (i.e. \qty{32}{\Hz}), we base it on a simple moving average of length $R_1 = 19$ where all coefficients $a_n = 1 / R_1$. The transfer function of a moving average is given by
\begin{equation}
    \tilde h_\mathrm{MA}(f) = \frac{1}{R_1} \frac{\sin(\pi f R_1 T_\mathrm{s})}{\sin(\pi f T_\mathrm{s})} ,
\end{equation}
which has zeros at the desired frequencies. Those are visible in the indigo line in the left panel of \fref{fig:design_cheap_filter}. We apply this filter five times to obtain sufficient attenuation in the vicinity of the nulls. The number of coefficients to represent an equivalent filter is $5 \cdot (R - 1) + 1 = 91$ as also stated in \tref{tab:filter_parameters}.

The remaining filters all decimate by a factor of two which simplifies their design. The sums in \eref{eq:psd_aliasing_individual} collapse to a single term as only a single frequency band has to be folded. We rely on type II \gls{fir} filters\footnote{\Gls{fir} type II filter have an odd number of coefficients with even symmetry. We restrict ourselves to this class of \gls{fir} filters as they have an integer sample group delay which can be trivially compensated.} and make heavy use of the Parks-McClellan algorithm which optimizes filter coefficients for a set of prescribed performance requirements. The Parks-McClellan algorithm find the filter coefficients $\{a_n\}_{n=0,\dots,N-1}$ that minimize the maximum (absolute) weighted error $|E(f)|$ over a given domain $U$. The weighted error is defined as 
\begin{equation}
    E(f) = W(f) \big(\tilde h(f) - D(f)\big) , \label{eq:weigthed_error_fir}
\end{equation}
where $D(f)$ denotes the desired frequency response of the filter. The actual frequency response $\tilde h(f)$ can be rewritten as a polynomial $P(x)$ in $x = \cos(2\pi f T_\mathrm{s})$ of degree $(N-1) / 2$. To force a unity response at DC ($f = 0$ or, equivalently, $x = 1$) and a vanishing response at the Nyquist rate ($f = f_\mathrm{s} / 2$ or, equivalently, $x = -1$) we redefine the polynomial as $P(x) = \big(\hat P(x) \cdot (x - 1) + 1/2\big) \cdot (x + 1)$ which transforms the error function (for brevity, now defined in terms of $x$) as
\begin{align}
    E(x) &= W(x) \cdot \Big(\big(\hat P(x) \cdot (x - 1) + 1 / 2\big) \cdot (x + 1) - D(x)\Big) , \\
    &= \underbrace{\vphantom{\bigg(}W(x) \cdot (x - 1) \cdot (x + 1)}_{\hat W(x)} \bigg(\hat P(x) - \underbrace{\vphantom{\bigg(}\frac{D(x) / (x + 1) - 1 / 2}{x - 1}}_{\hat D(x)}\bigg) .
\end{align}
The resulting polynomial $\hat P(x)$ of the optimization procedure has two degree less than $P(x)$. The filter coefficients $a_n$ are determined by computing the inverse Fourier transform of the resulting response $\tilde h(f) = P(\cos(2 \pi f T_\mathrm{s}))$ at distinct frequencies $f_n = n f_\mathrm{s} / N$ where $n = 0, \dots, N - 1$. The domain $U$ for a filter (decimating by a factor of two) running at $f_\mathrm{s}$ covers only the stop-band ranging from $f_\mathrm{s} / 2 - \qty{1}{Hz}$ up to the Nyquist rate $f_\mathrm{s} / 2$ to limit aliasing into the \gls{lisa} band. Within $U$ the desired response is $D(f) = 0$.

To ensure that the aliased power has a $\propto\! f$ dependency the weighting function takes the simple form
\begin{equation}
    W(f) = \frac{1}{f - f_\mathrm{s} / 2} .
\end{equation}
This choice of weights approaches infinity at the Nyquist rate which seemingly drives the weighted error function in \eref{eq:weigthed_error_fir} to infinity. However, we have made sure that $P(x) - D(x)$ vanishes at the Nyquist frequency which overall drives the weighed error to zero.

The resulting filter designs for FIR 2 to 4 are depicted in \fref{fig:design_cheap_filter} as the sand, green and cyan lines. All filters (including the CIC filter and FIR 1) perform adequately with respect to aliasing (left panel) as the aliased contributions (computed using left-hand side of \eref{eq:aliasing_requirement}) stay below the \qty{1}{\pico\m} reference curve. The right panel shows the (scaled) transfer function derivatives. We note that the final FIR filter (cyan) causes the total derivative of the whole filter chain to rise above the \qty{0.1}{\pico\m} reference curve for frequencies greater than \qty{0.3}{\Hz}.

To correct for the residual non-flatness of the total chain close to the upper edge of the \gls{lisa} band we introduce a compensation filter $\filter^+$ running at the final sampling rate of \qty{4}{\Hz} that lifts the response in band to flatten out the transfer function, or, in other words, reduce the magnitude of its derivative.

We determine the coefficients of the compensation filter by setting up a similar optimization problem as in \eref{eq:weigthed_error_fir} and solving it using the Parks-McClellan algorithm. In this case we require that the derivative of the combined transfer function of the full filter chain and the compensation filter should be close to zero in $U = [\qty{0}{\Hz}, \qty{1}{\Hz}]$. We define as the weighted error function,
\begin{align}
    E(f) &= W(f) \cdot \frac{\mathrm{d}}{\mathrm{d} f} \left(\tilde h_{\filter^+}(f) \cdot \tilde h_{\filter}(f)\right) \\
    &= W(f) \cdot \left(\tilde h'_{\filter^+}(f) \cdot \tilde h_{\filter}(f) - \left(-\tilde h_{\filter^+}(f) \cdot \tilde h'_{\filter}(f)\right)\right) \\
    \begin{split}
        &= \underbrace{W(f) \cdot \tilde h_{\filter}(f) \cdot \big(-2 \pi T_\mathrm{s} \cdot \sin(2 \pi f T_\mathrm{s})\big)\vphantom{\Bigg(}}_{\hat W(f)} \\
        &\hspace{10em}\cdot \Bigg(\underbrace{P'(\cos(2 \pi f T_\mathrm{s}))\vphantom{\Bigg(}}_{\hat P(\cos(2 \pi f T_\mathrm{s}))} - \underbrace{\frac{\tilde h_{\filter^+}(f) \cdot \tilde h'_{\filter}(f)}{2 \pi T_\mathrm{s} \cdot \sin(2 \pi f T_\mathrm{s}) \cdot \tilde h_{\filter}(f)}\vphantom{\Bigg)}}_{\hat D(f)}\Bigg) . \label{eq:weighted_error_compensation}
    \end{split}
\end{align}
Here, we make use of the fact that we can write $\tilde h'_{\filter^+}(f) = \frac{\mathrm{d}}{\mathrm{d}f} P(\cos(2 \pi f T_\mathrm{s}))$. As before, we identify the effective weight $\hat W(f)$, polynomial $\hat P(x)$ and desired response $\hat D(f)$.

Minimizing the maximum of the absolute weighted error in \eref{eq:weighted_error_compensation} is non-trivial. The desired response $\hat D(f)$ is dependent on the transfer function of the compensation filter we are solving for. To circumvent this problem we run the optimization procedure for multiple iterations using the previous result of $\tilde h_{\filter^+}(f)$ to calculate $\hat D(f)$ (starting from $\tilde h_{\filter^+}(f) = 1$ for the initial iteration). We find that the design converges after approximately ten iterations.

\begin{figure}
    \centering
    \includegraphics{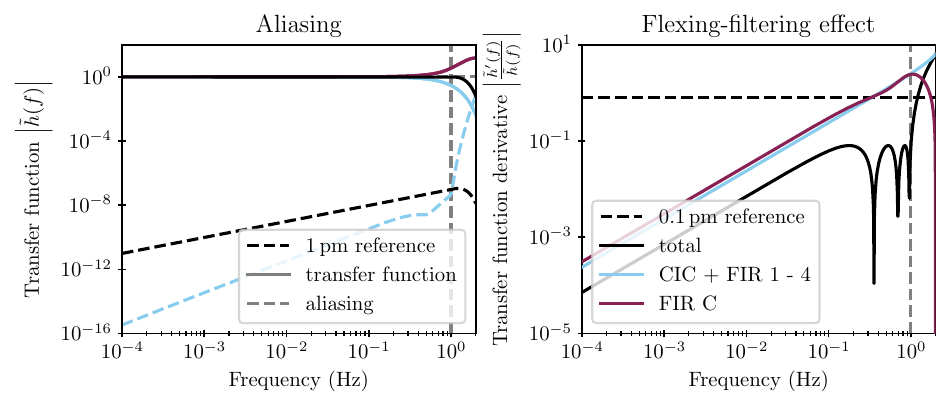}
    \caption{Frequency response of the overall filter chain including the CIC filter and FIR 1 - 4 (cyan) and compensation filter (wine). The plotted lines are analogous to \fref{fig:design_cheap_filter}. The transfer function of the combined response (black) appears much more flat in band and respects the \qty{0.1}{\pico\m} reference curve. We note that the transfer function derivative of the overall filter chain and the compensation filter have opposite sign (minus and plus, respectively). This is not apparent as we only plot the magnitude here.}
    \label{fig:design_flat_filter}
\end{figure}

The final design of the compensation filter and its performance is presented in \fref{fig:design_flat_filter}. The compensation filter (wine) is not causing any additional aliasing but shapes the aliased components from the previous stages. As desired it flattens out the response of the overall filter chain including the CIC filter and FIR 1 - 4 (cyan) where the maximum residual non-flatness appears constant. Most importantly the compensated response respects the \qty{0.1}{\pico\m} reference curve for the flexing-filtering effect.

\section{Flexing-Filtering Correction Design}
\label{app:correction_design}
The design procedure for the kernel function $k_\mathcal{H}(\tau)$ was introduced in \sref{sec:derivation}. Here we give an account of the algorithmic details of its implementation. Large parts of it are adopted from~\cite{Staab:2024nyo}.

For the sake of brevity we substitute $\xi = N T_\mathrm{s} f$ which lets us redefine $\tilde k_\mathcal{H}(\xi) = R(\xi) \cdot P(\xi^2)$ where $R(\xi)$ is an analytical function defined in~\cite{Staab:2024nyo} and $P(x)$ is a polynomial of degree $N - L$. The integer $L$ determines the smoothness of the kernel function $k_\mathcal{H}(\tau)$ at $\tau = \pm \frac{N T\mathrm{s}}{2}$. We choose $L = 2$ to yield a continuous (and as a side effect differentiable) kernel function. The transformed error function then reads
\begin{align}
    E(\xi) &= W(\xi) \cdot R(\xi) \bigg(P\left(\xi^2\right) - \frac{D(\xi)}{R(\xi)}\bigg) , \\
    &= \underbrace{\vphantom{\frac{D(\xi)}{\xi^2 R(\xi)}}W(\xi) \cdot \xi^2 \cdot R(\xi)}_{\hat W(\xi)} \bigg(\hat P\!\left(\xi^2\right) - \underbrace{\frac{D(\xi)}{\xi^2 \cdot R(\xi)}}_{\hat D(\xi)}\bigg) .
\end{align}
In the second line we set $P(\xi^2) = \hat P(\xi^2) \cdot \xi^2$. This ensures that the desired response is achieved at DC as this measure effectively forces $\tilde k_\mathcal{H}(0) = 0$. Note that the polynomial degree of $\hat P(x)$ is reduced by one compared to $P(x)$. Furthermore, we use a similar weighting function as in~\cite{Staab:2024nyo} to account for the ``red'' shape of the reference curve in frequency units. It is given by
\begin{equation}
    W(f) = 
    \begin{cases*}
        (f + f_\mathrm{min})^{-1} & if $0 \le f \le f_\mathrm{pass}$, \\
        10^3 \cdot f_\mathrm{pass}^{-1} \cdot (f / f_\mathrm{stop})^3 & else,
    \end{cases*} \label{eq:weightings}
\end{equation}
and features stronger weights for the stop-band 

The solution of this optimization problem is the polynomial $\hat P(x)$ that is evaluated to relate it back to the coefficients $a_n$ of the cosine-sum kernel in \eref{eq:cosine-sum}. This relation is given in~\cite{Staab:2024nyo} and is adapted for the choice $P(x) = \hat P(x) \cdot x$ as
\begin{equation}
    a_m = \frac{(-1)^m}{N T_\mathrm{s}} \cdot \frac{\hat P\!\left(m^2\right) \cdot m^2}{\prod_{\substack{n=0 \\ n \neq m}}^{N-1} m^2 - n^2} .
\end{equation}
The coefficients are then plugged into \eref{eq:cosine-sum} to evaluate the kernel function $k_\mathcal{H}(\tau)$ for the application of the operation $\mathcal{H}$ given in \eref{eq:discrete_convolution}.

\FloatBarrier

\section*{References}
\bibliographystyle{iopart-num}
\bibliography{refs}

\end{document}